# Bistable systems with Stochastic Noise: Virtues and Limits of effective Langevin equations for the Thermohaline Circulation strength

Valerio Lucarini♦ (1,2), D. Faranda (1), M. Willeit (3)

(1) Klimacampus, Institute of Meteorology, Hamburg Universität, Hamburg, Germany
(2) Department of Mathematics and Statistics, University of Reading, Reading, UK
(3) Potsdam Institute for Climate Impact Research, Potsdam, Germany

## Abstract

The understanding of the statistical properties and of the dynamics of multistable systems is gaining more and more importance in a vast variety of scientific fields. This is especially relevant for the investigation of the tipping points of complex systems. Sometimes, in order to understand the time series of given observables exhibiting bimodal distributions, simple one-dimensional Langevin models are fitted to reproduce the observed statistical properties, and used to investing-ate the projected dynamics of the observable. This is of great relevance for studying potential catastrophic changes in the properties of the underlying system or resonant behaviours like those related to stochastic resonance-like mechanisms. In this paper, we propose a framework for encasing this kind of studies and show, using simple box models of the oceanic circulation and choosing as observable the strength of the thermohaline circulation. We study the statistical properties of the transitions between the two modes of operation of the thermohaline circulation under symmetric boundary forcing and test their agreement with simplified one-dimensional phenomenological theories. We extend our analysis to include stochastic resonance-like amplification processes. We conclude that fitted one-dimensional Langevin models, when closely scrutinised, may result to be more ad-hoc than they seem, lacking robustness and/or well-posedness. They should be treated with care, more as an empiric descriptive tool than as methodology with predictive power.

♦ email: valerio.lucarini@zmaw.de

# Table of Contents

## 1. Introduction

An interesting property of many physical systems with several degrees of freedom is the presence of multiple equilibria (or, more in general, of a disconnected attractor) for a given choice of the parameters. In such a case, the system automatically does not obey ergodicity and its asymptotic state depends on what is the basin of attraction its initial condition belongs to. Among the many interesting properties of multi-stable systems is their possibility of featuring hysteretic behaviour: starting from an initial equilibrium $x = x_{in}$ realized for a given value of a parameter $P = P_{in}$ and increasing adiabatically the value of $P$ so that the system is always at equilibrium following $x = x(P)$, we may eventually encounter bifurcations leading the system to a new branch of equilibria $x = x'(P)$ such that, if we revert the direction of variation of P, we may end up to a different final stable state $x_{fin} = x'(P_{in}) \neq x(P_{in}) = x_{in}$. More generally, we can say that the history of the system determines which of the stable states is realized for a given choice of the parameters.

Whereas hysteretic behaviour has first been discussed in the context of magnetism, geophysical fluid dynamics offers some outstanding examples where multistability is of great relevance. Most notably, the problem which has probably attracted the greatest deal of interest is that of the stability properties of the thermohaline circulation (THC). Since paleoclimatic evidences suggest that the large scale circulation of the Atlantic Ocean presents at least two, qualitatively different, stable modes of operation (Boyle et al. 1987), theoretical and modellistic efforts have long been directed to understanding the mathematical properties of the circulation and the physical processes responsible for switching from one to the other stable mode and those responsible for ensuring the stability of either equilibrium.

Interestingly, it has been possible to construct very simple models of the THC (Stommel 1961, Rooth 1982) able to feature most of the desired properties, and models of higher degrees of complexity have basically confirmed the robustness of such properties of multistability, from simple two-dimensional convective equations models (Cessi and Young 1992, Vellinga 1996, Lucarini et al. 2005, 2007) to simplified climate models (Stocker and Wright 1991, Rahmstorf 1995, Stocker and Schmittner 1997). Whereas climate models of intermediate complexity now consistently represent the THC as a multistable system (Rahmstorf et al. 2005), results are not conclusive when full 3D climate models are considered (Stouffer and Manabe 2003, Scott et al. 2008). Nonetheless, recent simulations performed by Hawkins et al. (2011) in a full 3D climate model have successfully reproduced the kind of bistability properties shown by Rahmstorf et al. (2005). In the case of THC, the strength of the hydrological cycle plays the role of dominant parameter, whose variation can lead the system through bifurcations (Sijp and England 2006, 2011; Sijp et al. 2011). A detailed account of these analyses can be found in Rahmstorf (1995), Scott et al. (1999), and Titz et al. (2001, 2002). The matter is of great relevance for understanding climate variability and climate change because, if the system is close to a bifurcation point, small changes in the parameters value could have virtually irreversible effects, driving the climate system to qualitatively different steady states. As evidenced in many studies (see, e.g. Kuhlbrodt et al. 2007), a transition from the present state of the THC to a state featuring weaker meridional circulation would have very relevant climatic effects at regional and global scale, as the northward ocean heat transport in the Atlantic would be greatly reduced.

The potential of shut-off of the THC is considered a high impact climate risk - even if its likelihood for the present climate is considered very low (Rahmstorf 2006) – and the conditions under which such a transition can occur are probably the best example of a "climate tipping point" (Lenton et al. 2008).

Since we are dealing in principle with a 3D fluid with complex thermodynamical and dynamical properties, a lot of efforts have been directing at finding, using suitable scaling and simplified theoretical setting, an approximate one-dimensional ordinary differential equation equation of the form $\dot{q} = \mathcal{F}(q, P)$, where $q$ is the intensity of the THC and $P$ is a set of parameters of the system. Such an equation would be able to represent at least in a semi-quantitative way the evolution of the THC strength as a function of the strength and some parameters only and, by solving $\mathcal{F}(q, P) = 0$, provide the (in general) multiple equilibria corresponding to a specific choice of the set of parameters $P$. An excellent account of this methodology can be found in Dijkstra (2005). Note that, very recently, a related surrogate one-dimensional dynamics for a salinity indicator has been proposed to fit the output of a comprehensive climate model (Sijp et al. 2011).

The dynamics of multistable systems becomes rather interesting when stochastic forcing is considered. In the most basic case, such a forcing is represented in the form of additive white noise. Noise introduces on one side small scale variability around each of the stable equilibria of the system, and, on the other side, allows for jumps (large scale variability) driving the system across the boundaries separating the basin of attractions of the fixed points. See Friedlin and Wentzel (1998) for a detailed mathematical treatment of these problems. Since the landmark Hasselmann's (1976) contribution, it has become clearer and clearer in the climate science community that stochastic forcing components can be treated as quite reliable surrogates for high frequency processes not captured by the variables included in the climate model under consideration (Fraedrich 1978, Saltzman 2002). This has raised the interest in exploring whether transitions between stable modes of operation of the THC far from the actual tipping points could be triggered by noise, representing high-frequency (with respect to the ocean's time scale) atmospheric forcings, of sufficient amplitude (Cessi 1994, Monahan 2002). Along these lines, it has become especially tempting to interpret the dynamics of the THC in presence of noise as resulting from an effective Langevin equation of the form $dq = \mathcal{F}(q,P)dt + \varepsilon dW$ - where $dW$ is the increment of a Wiener process - as this opens the way to approaching the problem in terms of one-dimensional Fokker-Planck equation. Ditlevsen (1999) suggested the possibility of considering more general stochastic processes for accommodating the statistical properties of observational data.

The Langevin equation approach has also led various researchers to study whether the process of stochastic resonance (Gammaitoni et al. 1998) - basically noise-enhanced response amplification to periodic forcing - could explain the strength of the climate response (in terms of actual THC strength) in spite of the relative weakness of the Milankovic forcing. Note that stochastic resonance, which has enjoyed great success in fields ranging from microscopic physics to neurobiology and perception, was first proposed in a climatic context (Benzi et al. 1982, Nicolis 1982, Benzi 2010). Velez-Belchi et al. (2001) provided the first example of a simple THC box-model featuring stochastic resonance, and later Ganopolski and Rahmstorf (2002) observed a similar mechanism in action in a much more realistic climate model.

In this work we would like to examine critically the effectiveness and robustness of using one-dimensional Langevin equations to represent the dynamical and statistical properties of the THC strength resulting from models which feature more than one degree of freedom. In Section 2 we will show how to construct, from data, the form of the effective driving force and the intensity of noise, and propose tests for investigating the robustness of the approach. In Section 3 we describe the main properties of the simple box models of the oceanic circulation analysed in this paper. In Section 4 we test our methodology by studying, in various cases, whether it is possible to represent consistently the statistical properties of the THC strength resulting from the stochastic forcing of the models introduced in Section 3 using an one-dimensional SDEs. In Section 5 we further expand our analysis by testing whether the matching conditions for observing stochastic resonance are obeyed. In Section 6 we present our conclusions.

## 2. Theoretical Background

### 2.a Inverse modelling

Let's consider a one dimensional Langevin equation for the variable *x* of the form:

$$dx = \mathcal{F}(x)dt + \varepsilon dW, \qquad (1)$$

where $\mathcal{F}(x)$ is a smooth function of $x$ giving the drift term, $W$ is a standard Wiener process and $dW$ is its infinitesimal increment, so that $\varepsilon$ parameterises the strength of the stochastic forcing. As well known, the invariant probability density function (pdf) $\pi(x)$ can be written as:

$$\pi(x) = Ce^{-2\frac{V(x)}{\varepsilon^2}}, \qquad (2)$$

where $V(x)$ is the effective potential such that $dV(x)/dx = -\mathcal{F}(x)$ and $C$ is the normalisation constant. The local extrema of the potential correspond to the fixed points of the deterministic system obtained when $\varepsilon = 0$, and in particular its local minima (maxima) giving the stable (unstable) equilibria. Quite intuitively, in the stochastic case, the peaks of the invariant probability distribution correspond to the minima of the potential.

In the prototypical situation of a confining double well potential, where we refer to the position of the right and left minimum of $V(x)$ as $x_+$, $x_-$, and to local maximum as $x_0$, the two peaks of the $\pi(x)$ are separated by a dip corresponding to the local maximum of the potential, while for large positive and negative values of x the $\pi(x)$ approaches zero as $V(x)$ diverges for $|x| \to \infty$. The average rate of transition $r(+\to -)$ from the basin of attraction of $x_+$ to that of $x_-$ can be approximated using the Kramer's formula:

$$r(+\to -) = \frac{1}{2\pi}\sqrt{V''(x_+)V''(x_0)}e^{-2\frac{V(x_0)-V(x_+)}{\varepsilon^2}}, \qquad (3)$$

under the condition that the absolute value of the exponent is larger than one, so that $V(x_0) - V(x_+) \gtrsim \varepsilon^2/2$. This corresponds to the physical condition that the noise is moderate with respect to the depth of the potential well. The average rate of transition $r(-\to +)$ in the opposite direction can be obtained by exchanging the sign plus with the sign minus in the previous expression. The Kramers' formula basically expresses the general fact that at stationary state a detailed balance conditions applies.

Let's now consider the case where we observe a scalar output signal $y(\vec{x})$ generated by a stochastic or chaotic deterministic flow of the system $\vec{x}$ living, in general, in an N-dimensional phase space and let's assume that the empirical pdf $\pi(y)$ is bimodal, so that two peaks are found at $y = y_+, y = y_-$, separated by a local minimum at $y = y_0$. We wish to test the possibility of constructing a Langevin equation providing the minimal stochastic model for the observed process in the form:

$$dy = \mathcal{F}_{eff}(y)dt + \varepsilon_{eff}dW. \qquad (4)$$

Such a representation would bypass the details of the full dynamics of the system and its construction can be approached by imposing constraints based upon the populations of the basin of attractions of the two modes and upon the transition probability between such basins. Basically, this amounts to defining an effective projected dynamics.

The observation time of the variable $y$ must be long enough to allow for a robust estimate of the pdf and for observing many transitions between the two modes. From the empirical pdf of the considered observable $\pi(y)$ we derive the function $(y) = -\ln(\pi(y))$. In order to achieve compatibility with Eq. (2), we define

$$2V_{eff}(y)/\varepsilon_{eff}^2 = U(y) + const. \tag{5}$$

The function $U(y)$ contains information on both the effective potential of the system and on the effective intensity of the noise. Substituting the expression of $V_{eff}(y)$ in Eq. (3), we can write, e.g., the average rate $r(+\to -)$ as follows:

$$r(+\to -) \approx \frac{\varepsilon_{eff}^2}{4\pi}\sqrt{U''(y_+)U''(y_0)}\,e^{-(U(y_0)-U(y_+))} \tag{6}$$

By comparing this expression with the observed transition rate, we can finally find the actual value of $\varepsilon_{eff}$, because all the other terms can be computed from the time series of the $y$ observable. Assuming that in the region between the two minima and the local maximum the function $U(y)$ is smooth, we obtain the following expression, which is numerically more robust as the second derivatives disappear:

$$r(+\to -) \approx \frac{\sqrt{2}\varepsilon_{eff}^2}{\pi}\frac{(U(y_0)-U(y_+))}{(y_+ - y_0)^2}e^{-(U(y_0)-U(y_+))} = \frac{2\sqrt{2}}{\pi}\frac{(V(y_0)-V(y_+))}{(y_+ - y_0)^2}e^{-\left(2\frac{V(y_0)-V(y_+)}{\varepsilon_{eff}^2}\right)} \tag{7}$$

where $G(+\to -) = \sqrt{2}/\pi\,(U(y_0)-U(y_+))/(y_+-y_0)^2\,e^{-(U(y_0)-U(y_+))}$ is a factor depending only on the observed probability. Note that Eqs. (6)-(7) are valid under the condition that $U(y_0)-U(y_+) \gtrsim 1$ or, equivalently, that $\pi(y_+)/\pi(y_0) \gtrsim e$. By plugging the obtained value of $\varepsilon_{eff}$ into Eq. (5) we derive $V_{eff}(y)$. We can then reconstruct the effective Langevin equation in the form given in Eq. (4). Note that, in order to have a consistent picture, the same results must be obtained when using as benchmark the average rate $r(-\to +)$, because consistency between the two approaches relies on the general principle of detailed balance. Such an issue is not relevant in the special case when the pdf of the system is symmetric with respect to $y_0$.

## 2.b    Conditions for Robustness

Let's now consider an N-dimensional Langevin equation of the form:

$$dx_i = F_i(x_1, \ldots, x_N)dt + \varepsilon_{ij}dW_j \qquad (8)$$

where the $dW_j$ terms indicate increments of independent Wiener processes and the $F_i(x_1, \ldots, x_N)$ are the (generally nonlinear) drift terms. We can write the Langevin equation for the observable $y$ constructed as linear combination of the system variables $y = c_i x_i$ as follows:

$$dy = c_i \mathcal{F}_i(x_1, \ldots, x_N)dt + c_i \varepsilon_{ij}dW_j = c_i \mathcal{F}_i(x_1, \ldots, x_N)dt + \bar{\varepsilon}dW \qquad (9)$$

where, exploiting the independence of the N Wiener processes, we have $\bar{\varepsilon} = \sqrt{\sum_{j=1}^{N}(c_i \varepsilon_{ij})^2}$. If the deterministic part of system (9) features two stable equilibria $\vec{x}_+$ and $\vec{x}_-$, when stochastic noise is added, we will see hopping between the basins of attraction of these two points. When looking at the variable $y = c_i x_i$ as output of the system, we will see a bimodal distribution where the two peaks are centred at $y_+ = c_i x_{i,+}$ and $y_- = c_i x_{i,-}$, respectively, with a local minimum in-between situated at $y_0$. In such a case, as discussed above, we have a procedure to attempt to model the stochastic dynamics of the variable $y$ using a heuristic model in the form given in Eq. (4).

Comparing Eq. (4) and Eq. (9), we understand that the difference in the drift terms $c_i \mathcal{F}_i(x_1, \ldots, x_N) - \mathcal{F}_{eff}(y)$ describes the deterministic dynamics of the $y$ variable which cannot be parameterised in terms of the $y$ variable alone. Therefore, the ratio $\bar{\varepsilon}/\varepsilon_{eff}$ gives a measure of how appropriate is the goodness of the simplified one-dimensional model, being close to one when the dynamics of the variable $y$ is truly quasi one-dimensional. We may expect that if $y$ is a slow variable, which retains the long term memory of the system, the impact of the faster variables on its evolution can be effectively decomposed in a deterministic part, expressed in terms of $y$ alone, and in a stochastic part, which contributes to determining the value of $\bar{\varepsilon}$ (Saltzman 2002).

Since constructing an ad-hoc one-dimensional model from a time-series is typically possible by following the lines described above, we need to introduce some criteria to test the robustness of the approach we have undertaken. Obviously, as mentioned above, choosing a suitable variable $y$ will be crucial in ensuring the effectiveness of this one-dimensional parameterisation:

- As a first condition of robustness, we may ask that if we multiply the noise intensity of the true system by a factor $\alpha$, so that $\varepsilon_{ij} \to \alpha \varepsilon_{ij}$ in Eq. (8), we obtain that correspondingly $\varepsilon_{eff} \to \alpha \varepsilon_{eff}$ and the effective potential $V_{eff}(y)$ is not altered in Eq. (4).

- Another condition of robustness is that if in Eq. (9) we alter the noise matrix $\varepsilon_{ij}$ in such a way that $\bar{\varepsilon}$ is not altered in Eq. (9), we would like that, correspondingly, $\varepsilon_{eff}$ and $V_{eff}(y)$ are not altered in Eq. (4).

These conditions basically require that the reconstructed deterministic drift term is independent of the intensity of the noise and the reconstructed noise intensity scales linearly with the actual noise applied to the system, so that we can construct a relationship such as $\varepsilon_{eff}(\bar{\varepsilon}) \approx \gamma \bar{\varepsilon}$

## 3. Simple box models of the Thermohaline Circulation

### 3.a   Full Model

We consider the simple deterministic three-box model of the deep circulation of the Atlantic Ocean introduced by Rooth (1982) and thoroughly discussed in Scott et al. (1999) and Lucarini and Stone (2005a, 2005b). The model consists of a northern high-latitude box (box 1), a tropical box (box 2), and a southern high-latitude box (box 3). The volume of the two high-latitude boxes is the same and is 1/$V$ times the volume of the tropical box, where V is chosen to be equal to 2. The physical state of the box $i$ is described by its temperature $T_i$ and its salinity $S_i$; the box $i$ receives from the atmosphere the net flux of heat $H_i$ and the net flux of freshwater $F_i$; the freshwater fluxes globally sum up to 0, so that the average oceanic salinity is a conserved quantity of the system. The box $i$ is subjected to the oceanic advection of heat and salt from the upstream box through the THC, whose strength is $q$. The dynamics of the system is described by the evolution equation for the temperature and the salinity of each box. After a suitable procedure of non-dimensionalisation (Lucarini and Stone 2005a), we obtain the following final form for the temperature and salinity tendency equations for the three boxes:

$$\dot{T}_1 = \begin{cases} q(T_2 - T_1) + H_1 \\ |q|(T_3 - T_1) + H_1 \end{cases} \tag{10a}$$

$$\dot{T}_2 = \begin{cases} \frac{q}{V}(T_3 - T_2) + H_2 \\ \left|\frac{q}{V}\right|(T_1 - T_2) + H_2 \end{cases} \tag{10b}$$

$$\dot{T}_3 = \begin{cases} q(T_1 - T_3) + H_3, & q > 0 \\ |q|(T_2 - T_3) + H_3, & q \leq 0 \end{cases} \tag{10c}$$

$$\dot{S}_1 = \begin{cases} q(S_2 - S_1) - F_1 \\ |q|(S_3 - S_1) - F_1 \end{cases} \tag{10d}$$

$$\dot{S}_2 = -\frac{1}{V}(\dot{S}_1 + \dot{S}_3) \tag{10e}$$

$$\dot{S}_3 = \begin{cases} q(S_1 - S_3) - F_3, & q > 0 \\ |q|(S_2 - S_3) - F_3, & q \leq 0 \end{cases} \tag{10f}$$

where $q = k(\rho_1 - \rho_3)$ and $\rho_i = \rho_0(1 - \alpha T_i + \beta S_i)$, so that $q = k\big(\beta(S_1 - S_3) - \alpha(T_1 - T_3)\big)$, $\alpha$ and $\beta$ are the usual thermal and haline expansion coefficients, $\rho_0$ is a baseline density, and $k$ is the hydraulic constant controlling the water transport. Such a parameterisation was first introduced by Stommel for a hemispheric box model, whereas the approximate linear relationship between the density difference between the two high-latitude regions and the THC strength was confirmed by simplified yet realistic GCM simulations (Rahmstorf 1995, Scott et al. 2008).

The freshwater fluxes $F_i$ are considered given constants, with $F_2 = -(F_1 + F_3)/V$, so that $S_1 + VS_2 + S_3 = (V + 2)S_0$ at all times, where $S_0 = 35\ psu$ is a baseline salinity. Instead, the heat flux $H_i = \lambda(\bar{T}_i - T_i)$ is such that the box temperature is relaxed to a fixed target temperature $\bar{T}_i$ with the time constant $\lambda^{-1}$. Such a representation mimics the combined effect of radiative heat transfer and of a meridional heat transport. This also implies that the spatial average of ocean temperature $\tilde{T} = (T_1 + VT_2 + T_3)/(2 + V)$ obeys the evolution equation $\dot{\tilde{T}} = \lambda(\bar{\tilde{T}} - \tilde{T})$, where $\bar{\tilde{T}}$ is the spatial average of the target temperature, so that asymptotically (and, in practise, after few units of $\lambda^{-1}$) $\tilde{T}$ is a conserved quantity. Therefore, we practically have $\dot{T}_2 \approx -(\dot{T}_1 + \dot{T}_3)/V$. We usually have that the internal time scale of the system $q^{-1}$ is much larger than $\lambda^{-1}$, thus implying that the thermal relaxation is fast.

We now present some basic properties of the system (10a-f) following Scott et al. (1999). Starting from random initial conditions, since the system has symmetric boundary conditions, we have equal probability of falling into either the northern sinking state ($q > 0$) or the southern sinking state ($q < 0$). The water of the high latitude ocean box where downwelling occurs is warmer and more saline that than situated on the opposite side of the planet, since it receives advection from the warm and saline equatorial box. Since the haline contribution is stronger, the downwelling box is denser than the upwelling box, and determines the sign of $q$. Instead the thermal advection negative feedback dominates the haline advection positive feedback and determines the stability of the realised equilibrium. The buoyancy fluxes in the upwelling and downwelling boxes serve different purposes in determining the dynamics of the system. In fact, the strength of the circulation at equilibrium $q_{ref}$ depends only on the strength of the buoyancy fluxes in the upwelling box $H_{u,eq}$ and $F_u$:

$$|q_{ref}| = \sqrt{k(\alpha H_{u,eq} + \beta F_u)},$$

where the sign of $q_{eq}$ is positive if $u = 3$ and negative if $u = 1$. Instead, for a given value of $F_u$, the realised pattern of circulation is stable as long as $F_d \lesssim 3F_u$, which implies that $F_d = 3F_u$ a bifurcation leading to an instability of the system is found. Such an instability exchanges the role of the upwelling and downwelling boxes. Therefore, if, e.g. in the initial state $u = 3$ and $d = 1$, and $F_3$ is kept fixed, the system features bistability for the following range of values of $F_1$: $1/3\, F_3 \lesssim F_1 \lesssim 3F_3$. These approximate relations become exact in the limit of infinitely fast thermal relaxation.

Following Scott et al. (1999) and Lucarini and Stone (2005a), we select for the constants of the system the values $k = 1.5 \cdot 10^{-6} s^{-1}, \alpha = 1.5 \cdot 10^{-4} K^{-1}, \beta = 8.0 \cdot 10^{-4} (psu)^{-1}, \lambda = 1.3 \cdot 10^{-9} s^{-1}, V = 2$. When symmetric boundary conditions are considered with $\bar{T}_1 = \bar{T}_3 = 0°C, \bar{T}_3 = 30°C, F_1 = F_3 = 9 \cdot 10^{-11} psu\, s^{-1}$, we obtain at steady state $|q_{ref}| = 1.47 \cdot 10^{-11} s^{-1}$. The sign of $q$ depends uniquely on the initial conditions of the integration: we have 50% probability of ending up in either the northern or the southern downwelling state if random initial conditions are chosen. Since the internal time scale $|q_{ref}|^{-1} \approx 215\, y$ is much larger than the thermal time scale $\lambda^{-1} \approx 25\, y$, we conclude that the thermal relaxation is a fast process.

The physical value $\tilde{q}$ of the strength of the thermohaline circulation can be found from the normalised value above as $\tilde{q} = q V_{box,1}$, where $V_{box,1} = V_{box,3} = V_{box,2}/V = 1.1 \times 10^{17} m^3$ is the volume of either high-latitude box. Instead, the physical value of the net freshwater flux $\tilde{F}_i$ into box $i$ is obtained as $\tilde{F}_i = F_i V_{box,i}/S_0$; its intensive value per unit surface results to be $\tilde{F}_i = F_i V_{box,i}/A_{box,i} S_0 = F_i D_{box}/S_0$ where $D_{box} = 5000\, m$ is the common depth of the three oceanic boxes. Therefore, our base state features reasonable values for the net poleward transport of freshwater flux - about $2.8 \times 10^5\, m^3/s = 0.28\, Sv$, or, equivalently, $0.41\, m/y$ - and for the THC strength - about $1.55 \times 10^7\, m^3/s = 15.5\, Sv$.

As the THC strength $q$ describes the most relevant climatic signal and sets the slowest time scale of the system, we would like to discuss the possibility of defining, along the lines of what discussed in Section 2, an effective Langevin equation for $q$ able to mimic effectively its statistical properties resulting from the evolution of the full system (10a-f) modified by the inclusion of additive stochastic forcing representing high-frequency climatic variability (Velez-Belchi et al. 2001). In this direction, we first introduce a simplified box model for the density-driven oceanic circulation.

### 3.b    Simplified Model

A simplified version of the model given in Eqs. (10a-f) can be derived by assuming that the thermal restoring constant $\lambda \to \infty$ so that the time scale of the feedback $\lambda^{-1} \to 0$. Thus, the temperatures of the three boxes are such that at all times $T_i = \bar{T}_i$, so that we obtain a reduced dynamical system with only 2 degrees of freedom (d.o.f.):

$$\dot{S}_1 = \begin{cases} q(2S_0 - 3/2\, S_1 - 1/2\, S_3) - F_1 & q > 0 \\ |q|(S_3 - S_1) - F_1 & q \leq 0 \end{cases} \qquad (11a)$$

$$\dot{S}_3 = \begin{cases} q(S_1 - S_3) - F_3, & q > 0 \\ |q|(2S_0 - 3/2\, S_3 - 1/2\, S_1) - F_3, & q \leq 0 \end{cases} \quad (11b)$$

where the THC strength can be written as $q = k\beta(S_3 - S_1)$. Note that, since the system (11a-b) has been obtained by performing a singular perturbation to (10a-10f), we need to renormalize the value of the hydraulic constant $k$ in order to obtain $|q_{ref}| = 1.47 \cdot 10^{-11} s^{-1}$ at steady state when choosing $F_1 = F_3 = 9 \cdot 10^{-11} psu\, s^{-1}$ as above. The resulting value is $k = 3.0 \cdot 10^{-7} s^{-1}$, with $|q_{ref}| = \sqrt{k\beta F_u} = \sqrt{k\beta F_1}$. The same physical scaling describe above apply here. Such a simplified model retains the most relevant elements of the dynamics of the full model, even if the thermal dynamical feedbacks (Scott et al. 1999) are missing.

## 4. Numerical Experiments: Fitting the dynamics from the Population and the Transition rates

### 4.a  Symmetric Forcing to the simplified model

We now modify the dynamical system (11a-b) by including additive noise in both the evolution equations for both $S_3$ and $S_1$ so that we obtain the following system of stochastic differential equations in the Ito form:

$$dS_1 = \begin{cases} q(2S_0 - 3/2\, S_1 - 1/2\, S_3)dt - F_1 dt + \varepsilon_1 dW_1, & q > 0 \\ |q|(S_3 - S_1)dt - F_1 dt + \varepsilon_1 dW_1 & q \leq 0 \end{cases} \quad (12a)$$

$$dS_3 = \begin{cases} q(S_1 - S_3)dt - F_3 dt + \varepsilon_3 dW_3, & q > 0 \\ |q|(2S_0 - 3/2\, S_3 - 1/2\, S_1)dt - F_3 dt + \varepsilon_3 dW_3, & q \leq 0 \end{cases} \quad (12b)$$

where $dW_{1,3}$ are the increments of two independent Wiener processes. Note that Ditlevsen (1999) proposed the possibility of considering more general noise processes to explain the THC dynamics. Hereby, we stick to the more usual white noise case.

We then perform a set of experiments by integrating the stochastic differential equations (10a-b) using the numerical scheme proposed in Mannella and Palleschi (1989) for values of $\varepsilon_1 = \varepsilon_3 = \varepsilon$ ranging from $3.6 \times 10^{-10}\, psu\, s^{-1/2}$ to $6.2 \times 10^{-10}\, psu\, s^{-1/2}$. This corresponds to a range of noise strength for the physical freshwater flux of $1.12 \times 10^6\, m^3\, s^{-1/2}$ to $1.96 \times 10^6\, m^3\, s^{-1/2}$, or, equivalently, in intensive terms and referring to years, to a range between $1.69\, m\, y^{-1/2}$ and $2.96\, m\, y^{-1/2}$. In more concrete terms, we are exploring stochastic perturbations to the freshwater flux whose variability (standard deviation), over the characteristic internal time scale $|q_{ref}|^{-1} \approx 215\, y$, range between 27% and 47% of the baseline value $F_1 = F_3$. Results are presented for sets of 100 ensemble members for each value of $\varepsilon$, with each integration lasting $10^6 \ y$. The chosen time step is $1 \ y$.

We wish to study the possibility of defining up to a good degree of precision a consistent stochastic dynamics for the THC strength $q$ involving only $q$ itself and noise. Following the procedure outlined in Section 2, for each value of $\varepsilon_1 = \varepsilon_3 = \varepsilon$, where we have on purpose kept the system's parameters invariant with respect to exchanging the box 1 and the box 3, we attempt the derivation of the deterministic drift term and the stochastic noise defining the effective Langevin equation for the THC strength:

$$dq = F_{eff}(q,\varepsilon)dt + \varepsilon_{eff}(\varepsilon)dW, \qquad (13)$$

where our notation accommodates for a noise-dependent effective drift term, which corresponds to an efficient potential $V_{eff}(q,\varepsilon)$, such that $F_{eff}(q,\varepsilon) = -dV_{eff}(q,\varepsilon)/dy$. The pdfs $\pi(q)$ feature a very strong dependence on the intensity of the noise, with, as expected, higher noise intensity associated to flatter distributions (Fig. 1). We then derive the normalised potential $U(q,\varepsilon) = -\ln(\pi(q,\varepsilon))$ (Fig. 2). By matching the observed hopping rate $r(+\to -) = r(-\to +)$ (Fig. 3a) with the right hand side of the formula given in Eq. (7) - in Fig. 3b we present the values of the factor $(+\to -)$ - we derive for each value of $\varepsilon$ the corresponding value of $\varepsilon_{eff}$. Note that for each value of the noise we use only the observed difference between the value of $U(q,\varepsilon)$ evaluated in $q = 0$ and in $q = |q_{ref}|$ and the value of $|q_{ref}|$, and the upper bound of $\varepsilon$ has been chosen so that $U(q_0,\varepsilon) - U(q_+,\varepsilon) = U(q_0,\varepsilon) - U(q_-,\varepsilon) \gtrsim 1.5$. As shown in Fig. 3c, we obtain that up to a high degree of precision $\varepsilon_{eff} \approx \gamma\sqrt{2}k\beta\varepsilon = \gamma\bar{\varepsilon}$, where $\gamma \approx 1$ for all values of $\varepsilon$. Moreover, also is in agreement with our expectations given at the end of Section 2, we have that $V_{eff}(q,\varepsilon) \approx V_{eff}(q)$, so that the effect of adding noise does not impact the deterministic drift term, or, in other terms, a deterministic dynamics is well defined. Figure 4 shows that for all values of noise the obtained effective potentials collapse into a single universal function, apart from an additive constant of no physical significance.

Our experimental procedure has shown quite convincingly that we can reduce the time evolution of the THC strength to a single SDE. We wish now to investigate how to derive analytically the drift and the noise term in Eq. (13) and an expression for the hopping rate $r(+\to -)$. Rewriting the system (10a-b) with respect to the new variables $q = k\beta(S_1 - S_3) = k(\rho_1 - \rho_3)$ and $Q = k(\rho_1 + \rho_3)$, we obtain the following coupled evolution equations:

$$dq = q(2k\tilde{\rho} - 3/2\ q^2/|q| - Q)dt + \varepsilon_q dW_q \qquad (14a)$$

$$dQ = q(2k\tilde{\rho} + 1/2\ q^2/|q| - Q)dt - k\beta(F_1 + F_3)dt + \varepsilon_Q dW_Q \qquad (14b)$$

where $\tilde{\rho} = \rho_0\big(1 - \alpha\,(\tilde{T}_1 + 2\tilde{T}_2 + \tilde{T}_3)/4 + \beta S_0\big)$ is the average density of the system, we have that, following Eq. (7), $\varepsilon_q = \varepsilon_Q = k\beta\sqrt{\varepsilon_1^2 + \varepsilon_3^2}$, and $dW_q$, $dW_Q$ are increments of Wiener processes. Note that the drift terms in both Eq. (11) is odd with respect to the $q \to -q$ transformation and the SDE for the THC strength is in the form of Eq. (9), with $\varepsilon_q = \bar{\varepsilon}$. We assume that the system spends most of its time near the two deterministic equilibria with $q = \mp|q_{ref}|$, and that the $Q$ is a slow variable, so that its deterministic drift term $\approx 0$. Assuming that the random forcing on $Q$ has little impact on $q$, we derive the following approximate SDE for the evolution of the THC strength:

$$dq \approx -2q(q^2/|q| - |q_{ref}|)dt + \varepsilon_q dW_q, \tag{15}$$

where, in the drift term is odd with respect to parity and is independent of the noise strength. The corresponding effective potential $V_{eff}(q)$ is independent of $\varepsilon$ and can be written as $V_{eff}(q) = -q^2|q_{ref}| + 2/3\, q^2|q| + const$. Note that this is a not a quartic symmetric potential but has the same parity properties and is twice differentiable everywhere. As shown in Fig. 4, this functional form closely approximates the experimental findings previously described, with discrepancies where the probability density is exponentially vanishing and small deviations also for $q \approx 0$ (where the density is also low). Moreover, if $\varepsilon_1 = \varepsilon_3$, we obtain that $\varepsilon_q = \bar{\varepsilon} = \sqrt{2}k\beta\varepsilon_1 \approx \varepsilon_{eff}$, so that the agreement between our experimental and theoretical findings for both the deterministic and stochastic part of the dynamics is quite satisfactory. This suggests that in the experimental setting of Eqs. (12a-b) it is possible to project very efficiently the dynamics of the system on the variable $q$ alone, which seems to capture well the slow manifold (Saltzman 2002) of the system. Using Eq. (15), we obtain the following approximate expression for the average rate of transition $r(+\to -)$ between the northern sinking and the southern sinking state

$$r(+\to -) \approx \frac{2\sqrt{2}}{3\pi}|q_{ref}|e^{-\frac{|q_{ref}|^3}{3(k\beta\varepsilon)^2}} = \frac{2\sqrt{2}}{3\pi}\sqrt{k\beta F_1}\,e^{-\frac{(F_1)^{3/2}}{3\sqrt{k\beta}\varepsilon^2}} = r(-\to +), \tag{16}$$

where the last identity is due to the symmetry of the potential. This formula provides rates in excellent agreement with the outputs of the numerical simulations, as can be seen by comparing the red and the black line in Fig. 3a.

### 4.b    Asymmetric forcing to the simplified model

The obtained results suggest that the simplified model of the THC with only 2 degrees of freedom allows for a robust treatment of the one-dimensional stochastic dynamics of the THC strength. Nonetheless, in the previous set of experiments we have only verified the first condition for the robustness of the one-dimensional representation (well-posedness for linear scaling on the forcings).

In this section, we wish to test how the system behaves when, following Eq. (9), we change the noise matrix $\varepsilon_{ij}$ in such a way that the term noise strength $\bar{\varepsilon}$ for the considered observable is not altered. Therefore, we perform a new set of experiments, where the stochastic forcing is exerted only in one of the two boxes, e.g. box 1, so that in Eq. (12a-b) we set $\varepsilon_3 = 0$ and $\varepsilon_1 = \sqrt{2}\varepsilon$ for each value of $\varepsilon$ considered in the previous set of experiments. This choice guarantees that we have exact correspondence for the values of $\varepsilon_q = \bar{\varepsilon} = k\beta\sqrt{\varepsilon_1^2 + \varepsilon_3^2} = \sqrt{2}k\beta\varepsilon$, so that Eq. (9) looks exactly the same as in the previous set of experiments. Unfortunately, as shown in Fig. 5, following the procedure described in Section 2 we obtain for all values of $\varepsilon$ an asymmetric probability distribution function, with the northern sinking equilibrium being the most probable state. The prominence of the $q > 0$ conditions become stronger as we consider weaker intensities for the noise. Therefore, the proposed stochastic modelling is not as robust as one could have guessed.

At a second thought the presence of asymmetry in this case becomes clearer. In this case the two sinking states undergo different forcing, because when $q > 0$ the stochastic forcing is exerted only in the box where downwelling occurs, whereas when $q \leq 0$ the stochastic forcing impacts only the box where upwelling occurs. Since, as explained at the beginning of Section 3 and discussed in Lucarini and Stone (2005a), the impact of changing freshwater fluxes is different in terms of destabilising the system depending whether the forcing is applied in the box where downwelling or upwelling occurs, the two states are not equivalent anymore. In the previous set of experiments, as opposed to that, both boxes were equally (in a statistical sense) stochastically forced, and so that the northern and southern sinking states had equivalent forcings at all times. The more formal mathematical reason why the statistical properties of $q$ are different in the two sets of experiments even if Eq. (9) is apparently the same can be traced to the differences in the correlation between the stochastic forcings to $q$ and $Q$ - compare Eqs. (14a-b). In the case of symmetric stochastic forcing in the freshwater fluxes into the two boxes with $\varepsilon_1 = \varepsilon_3$, , it is easy to see that the increments to the Wiener processes $dW_q$ and $dW_Q$ are not correlated, whereas when $\varepsilon_3 = 0$ the two quantities $dW_q$ and $dW_Q$ are identical so that their correlation is unitary.

We wish now to test whether, in such an asymmetric setting of forcings, the pdfs of the THC strength scale with the intensity of the noise in such a way to allow the possibility of defining consistently an effective potential $V_{eff}(q, \varepsilon)$ driving the deterministic part of the one-dimensional stochastic evolution of the THC strength. Such an effective potential would, unavoidably, be different from the one derived in the previous set of experiments. If we are able to define such an effective potential, we can deduce that each choice of the correlation matrix for the noise in the full system determines a specific projected effective deterministic dynamics.

We follow the procedure described in Section 2, and for all the chosen values of $\varepsilon$ we have that $U(q_0,\varepsilon) - U(q_-,\varepsilon) \gtrsim 1.5$ and $U(q_0,\varepsilon) - U(q_+,\varepsilon) \gtrsim 1.5$. In Fig. 6a we present the hopping rates $r(+\to -)$ (blue) and $r(-\to +)$ (red). We see that both values increase monotonically with $\varepsilon$, as stronger noise favours transitions. Note that, quite unexpectedly, for $\varepsilon > 7 \times 10^{-10} psu\, s^{-1/2}$ (beyond the range where our full analysis is performed, not shown), $r(+\to -)$ becomes bigger. By simple population algebra, this implies that the fraction of states with $q < 0$ is larger than $1/2$, even if the most probable state is given, in all cases, by $q = q_{ref} > 0$. In Fig. 6b we plot the factors $G(+\to -)$ and $G(-\to +)$. By matching the rate of transition with the corresponding $G$ factor, one obtains for each value of $\varepsilon$ the effective noise intensity $\varepsilon_{eff}$ such that the probability distribution of the states and the transitions statistics are compatible. As mentioned before, in order to have a consistent picture, we need to obtain the same value of $\varepsilon_{eff}$ by either using the $+\to -$ or the $+\to -$ path. In Fig. 6c (compare with Fig. 3c) we show that $\varepsilon_{eff}$ does not scale linearly with $\varepsilon$ (or, equivalently, $\gamma$ is not a constant) as instead found in the previous case, and that there is no consistency between its value as obtained using the statistics of the $+\to -$ and $-\to +$ transitions. Therefore, we cannot reconstruct a well-defined effective potential $V_{eff}(q,\varepsilon)$, so that a consistent one-dimensional SDE representation of the dynamics and statistics of the THC strength is not possible here. This can be the case if noise can effectively activate non-trivial dynamical processes allowing for a transition between the neighbourhoods of the two steady states with $q = \pm|q_{ref}|$, where by non-trivial we mean that they cannot be represented even approximately as a function of $q$ only.

### 4.c  Symmetric Forcing to the Full System

We now revert to the full system described at the beginning of the present section. By adding stochastic perturbations to the freshwater fluxes in a similar fashion as in Eq. (11a-b), so that $F_j \to F_j - \varepsilon_j\, dW_j/dt$ with $j = 1,3$ in Eq. (10a-f) we obtain the following Langevin equations for the variables $q = k(\rho_1 - \rho_3)$ and $Q = k(\rho_1 + \rho_3)$:

$$dq = q(2\tilde{q} - 3/2\, q^2/|q| - Q)dt + k\alpha\lambda(T_1 - T_3)dt + \varepsilon_q dW_q, \qquad (17a)$$

$$dQ = q(2k\tilde{\rho} + 1/2\, q^2/|q| - Q)dt - \beta(F_1 + F_3)dt + k\alpha\lambda(\tilde{T}_1 - T_1 + \tilde{T}_3 - T_3)dt + \varepsilon_Q dW_Q, \quad (17b)$$

where the same notation as in Eqs. (14a-b) has been used. Note that, as opposed to Eq. (11), the deterministic drift term is not odd with respect to the $q \to -q$ transformation, since in this case explicit temperature dependent terms are present, so that a negative parity is realised only when the signs of both $T_1 - T_3$ and $S_1 - S_3$ are changed. Note that, following the same derivation as in the case of the system with 2 d.o.f. and assuming that $Q$ is a slow variable, we end up writing the same approximate autonomous Eq. (15) for the THC strength under the hypothesis that the system spends most time near the two deterministic fixed points $q = \pm |q_{eq}|$. This suggests that also in this case we might empirically derive a well-defined effective potential $V_{eff}(q)$ analogous to the one obtained for the 2 d.o.f. model if considering stochastic forcing acting on both boxes 1 and 3.

Therefore, we follow the analysis of the previous subsection, and concentrate to stochastic perturbations to the freshwater flux having the same strength $\varepsilon_1 = \varepsilon_3 = \varepsilon$ in both hemispheres. We first observe that, as anticipated, for a given value of $\varepsilon$ the distribution of the THC strength is flatter than in the case of the 2 d.o.f model (Fig. 7). In order to obtain a pdf analogous to what obtained in the 2 d.o.f. case, in the full model we need to consider a stochastic forcing smaller by about 25%. As before, we select values of $\varepsilon$ such that $U(q_0, \varepsilon) - U(q_+, \varepsilon) = U(q_0, \varepsilon) - U(q_-, \varepsilon) \gtrsim 1.5$ in order to be able to use Kramers' formula as a constraint to check the consistency of our data with the one-dimensional Langevin model. In Fig. 8a we report the hopping rates as a function of the intensity of the noise. The behaviour is qualitatively analogous to what shown in Fig. 3a for the 2 d.o.f. model, but, in agreement with the discussion above, the hopping rate between the $q > 0$ and the $q < 0$ states are consistently higher for the full model when the same stochastic forcing is considered. In Fig. 8b we present the factor $G(+\to -) = G(-\to +)$ introduced in Eq. (7), which depends uniquely on the ratio between the probability density at the two maxima and at the local minimum for $q = 0$. Fig. 8c, similarly to Fig. 3c, shows the proportionality constant $\gamma$ between the value of $\varepsilon_{eff}$ and the value of $\bar{\varepsilon} = \sqrt{2}k\beta\varepsilon$. The parameter $\gamma$ should be close to unity in the case the projection of the dynamics on $q$ is "trivial" (as in the case of the 2 d.o.f. model with symmetric stochastic forcing), and, more importantly, $\gamma$ should be approximately independent of $\varepsilon$. In the case analysed here both conditions are not satisfied. The first issue points to the fact that we need to renormalize the constants when developing a lower dimensional projected dynamics (which is exactly what we did when constructing the 2 d.o.f. model from the full model). In fact, if we consider as effective $k$ the one considered for the 2 d.o.f. model, the values of $\gamma$ are relatively close to 1 (check the scale on the right hand side of Fig. 8c). More critical is the presence of a nonlinear relation between $\varepsilon_{eff}$ and $\varepsilon$. See below for an interpretation.

When reconstructing from $U(q, \varepsilon)$ the actual effective potential $V_{eff}(q, \varepsilon)$ using Eq. (5), we obtain that the effective potential is a function of $q$ only, so that $V_{eff}(q, \varepsilon) = V_{eff}(q)$ (see Fig. 9) so that it is possible to disentangle completely the drift term from the stochastic forcing. Moreover, such potential is very similar to what obtained in the reduced model with 2 d.o.f., as can be seen by comparing Fig. 4 and Fig. 9. The height of the potential barrier between the two minima is slightly lower in the case of the full model analysed here, in agreement with the argument of the destabilising feedback due to the thermal restoring process (Lucarini and Stone 2005a). This can be explained as follows: since perturbations in the value of $q - |q_{eq}|$ are positively correlated to perturbations $T_1 - T_3$ thanks to advection, the contribution $k\alpha\lambda(T_1 - T_3)$ in Eq. (17a) weakens the force driving the system towards the nearby deterministic fixed point, thus enhancing its instability.

The good agreement between Figs. 4 and 9 implies that the deterministic dynamics of the THC strength is robustly consistent between the full and reduced model, as partly anticipated above. The main difference between the two is that in the full model the nonlinear feedbacks connecting the variables of the system change in a nontrivial, nonlinear way the effective surrogate noise acting on the $q$ variable taken as independent.

## 5. Numerical experiments: Stochastic Resonance

Stochastic resonance (Benzi et al. 1982; Nicolis 1982; Gammaitoni et al. 1998) is an exceedingly interesting process whereby noise amplifies the response of the system at the same frequency of a periodic forcing. Typically, it is realised when we consider a Langevin equation of the form:

$$dx = F(x)dt + A\sin(\omega t + \phi)dt + \varepsilon dW \tag{18}$$

where the drift term $F(x) = -dV(x)/dx$ derives from a (symmetric, but necessarily so) potential $V(x)$ with a double well-structure like those considered in this study. We can associate the time dependent drift $G(x,t) = F(x) + A\sin(\omega t + \phi)$ to a time dependent potential $W(x,t) = -\int dx F(x) - Ax\sin(\omega t + \phi) = V(x) - Ax\sin(\omega t + \phi)$. The periodic forcing modulates the bistable system, so that one stable state corresponding to one of the two minima results to be less stable than the other every half a period of the forcing, so that every period the populations of the neighbourhoods of these points are, alternatively, exponentially increased and suppressed by a factor $exp(\pm 2A(|x_+ - x_0|)/\varepsilon^2)$. As accurately discussed in Gammaitoni et al. (1998), when we tune the noise intensity $\varepsilon$ so that the inverse of the hopping rate given in Eq. (7) for the unperturbed system is approximately equal to half of the period $\pi/\omega$, the system is in a condition where there is maximum probability for leaving the less stable state into the more stable one, before, randomly, the system switches back. In this case, the system attains a high degree of synchronisation with the input periodic signal, so that the output is basically a square wave with constant phase difference with the sinusoidal forcing.

Interestingly, even if the stochastic resonance is apparently a highly nonlinear process, it can be accurately described using linear response theory, whereby one studies the amplitude of the output signal at the same frequency of the periodic forcing for various values of the noise. Such amplitude, in the weak field limit, is proportional to the amplitude of the periodic forcing $A$ (Gammaitoni et a. 1998).

Along these lines, we consider the 2 d.o.f. model, and take into account a periodic modulation to the freshwater fluxes, so that $F_1 \to F_1 + \phi \Delta F sin(\omega t)$ and $F_3 \to F_3 + \psi \Delta F sin(\omega t)$, with symmetric background forcing $F_1 = F_3$. Assuming that the stochastic forcing acts with equal strength $\varepsilon_1 = \varepsilon_3 = \varepsilon$ on both boxes 1 and 3, as analysed in subsection 4.a, we obtain that Eq. (15), which satisfactorily describes the one-dimensional stochastic dynamics of $q$, is modified as follows:

$$dq \approx -2q(q^2/|q| - |q_{ref}|)dt + k\beta(\phi - \psi)\Delta F sin(\omega t) + \varepsilon_q dW_q, \qquad (19)$$

which is exactly in the form of Eq. (15). As we see, for a given value of $\Delta F$, the strength of the periodic forcing to $q$ depends only on the absolute value of the difference $(\phi - \psi)$, and not separately on the values of $\phi$ and $\psi$. If the dynamics of $q$ is accurately described by a one-dimensional Langevin equation, we expect to be able to observe the process of stochastic resonance when $\omega$ and $\varepsilon_q$ are suitably matched.

In order to test this, we set the period $\omega = \omega_0 = 2\pi/19000\ y$ - where the period of 19000 years has been chosen because it is long compared to the internal time scale $|q_{ref}|^{-1} \approx 215\ y$ and has paleoclimatic relevance in conjunction to Milankovitch theory (Veleze-Bechi et al. 2001; Saltzman 2002) - choose a moderate value for the amplitude of the sinuisodal modulation $\Delta F = 9 \times 10^{-12} s^{-2}$ and study the amplitude of the $\omega_0$ frequency component of the times series of $q$ as a function of $\varepsilon$, and create an ensemble of 100 members for each considered setting. We can state that the phenomenology of stochastic resonance is well reproduced if a) we find the characteristic peak for the response in the vicinity of a value of $\varepsilon$ such that the corresponding hopping rate for the unperturbed system given in Fig. 3a is close to $\pi/\omega_0$, and b) such response depends, for all values of $\varepsilon$, on $(\phi - \psi)$ only. We refer to the scenario where $\phi = -\psi = 1/2$ as case 1, and the scenario where $\phi = 1, \psi = 0$ as case 2. Note that the case $\phi = 0, \psi = -1$ is identical to case 2 by symmetry. The results are reported in Fig. 10, with the black line corresponding to case 1 and the red line corresponding to case 2. The obtained curves for the amplitude of the response agree very accurately, especially considering the rather small uncertainty, and feature exactly the right shape as presented in Gammaitoni et al. (1998). We observe a relatively broad resonance for values of noise comparable to those inducing in the unperturbed system transitions with average rate similar to the semiperiod of the forcing. Finding quite accurately the signature of stochastic resonance is a further proof that in the special setting of the unperturbed system considered here the dynamics of $q$ is indeed quasi-one dimensional.

We want to contrast this positive outcome with what one obtains by adding periodic perturbations of the same form as above to an "unperturbed" state featuring stochastic forcing acting on box 1 only, described in subsection 4.b. We choose exactly the same forcing parameters as above and repeat the experiments using the same ensemble size. We refer to the scenario where $\phi = -\psi = 1/2$ as case 3, and the scenario where $\phi = 1, \psi = 0$ as case 4, and to the scenario where $\phi = 0, \psi = -1$ as case 5. Note that here case 4 and case 5 are not equivalent. We obtain (see Fig. 10) that in the three cases, whereas we obtain qualitatively and quantitatively analogous results for the normalised amplitude of the response of the output at the same frequency of the forcing, the curves are distinct with high statistical significance and differ also from what obtained for cases 1 and 2. Such discrepancy would not be possible if the dynamics of $q$ were accurately described with a one-dimensional effective potential plus stochastic noise plus periodic forcing. The fact that the three curves 3, 4, and 5 are not superimposed (and disagree with 1 and 2) further supports the fact that the dynamics of $q$ is not trivially quasi-one dimensional. Nonetheless, an obvious process of resonance (not strictly one-dimensional) is obviously still in place.

## 6. Conclusions

In this work we re-examined the classic problem of trying to reconstruct the effective stochastic dynamics of an observable from its time series in the special case of a clearly bimodal empirical probability density function. This issue is especially relevant in climatic and paleoclimatic research, where it is very tempting to try to deduce the large scale, qualitative properties of the climate system, their modulation with time, the potential presence of tipping points through the observations of long time series of proxy data (Livina and Lenton 2007; Livina et al. 2009). Furthermore, since amplifying mechanisms such as stochastic resonance have been proposed to explain enhanced low frequency variability of the oceanic circulation as a result of slow modulations of some parameters (Ganopolski and Rahmstorf 2002), such empirically reconstructed statistical/dynamical properties may be interpreted as starting points to deduce special "sensitivities" of the climate system. Therefore, an important question is to understand how accurate and robust these procedures of reconstruction are.

From this work it is apparent that from the statistical properties of the time series of an observable featuring a symmetric bimodal pdf it is relatively easy to construct the corresponding one-dimensional Langevin equation by determining the drift term and the intensity of the white noise by basically imposing a self-consistent population dynamics. Assuming that the observable is a function of the phase space variables of a stochastic dynamical systems, it is an obvious temptation to interpret the obtained equation as the description of the projected dynamics for the observable, where the impact of the other (in general, many) neglected degrees of freedom of the system contributes to defining the effective deterministic dynamics and to creating a surrogate white noise term. Nonetheless, if the pdf of the observable is not symmetric, the possibility of constructing a meaningful surrogate stochastic dynamics relies on the fact that one should be able to describe consistently the hopping process between the two attraction basins.

In this work we have considered two very simple box models of the oceanic circulation (Rooth 1982, Scott et al. 1999, Lucarini and Stone 2005), comprising two high latitude and a low latitude boxes with time-dependent temperature and salinity as testbeds for these methodologies. These models are able to reproduce the bistability properties of the thermohaline circulation, by featuring two possible asymmetric circulations (one mirror image of the other) in presence of symmetric external heat and freshwater forcings. In both models the circulation strength is parameterised as proportional to the difference between the densities of the two high-latitude boxes. The simpler 2 d.o.f. model is suitably derived from the full, 5 d.o.f. model by imposing a fixed temperature for the boxes.

We first impose stochastic forcing of the same intensity on the freshwater forcings to the two high-latitude boxes and observe that the resulting pdf of the thermohaline circulation strength is bimodal and symmetric. More importantly, for both models the dynamics of $q$ can be accurately described with a Langevin equation with a drift term derived from a one-dimensional effective potential plus stochastic noise. An excellent approximation to the true dynamics (as well as to the hopping rates) can be obtained in an explicit form by imposing that the sum of the densities of the two high-latitude boxes is a slow variable. The main difference between the two is that in the full model the nonlinear feedbacks acting on the variable we are neglecting alter in a nontrivial, nonlinear way the effective surrogate noise acting on the $q$ variable. In other terms, in the case of the full model a careful tuning of the noise allows for taking care very accurately – in a statistical sense - of the effect of all the variables we are neglecting. Our results are obtained for a specific value of the background freshwater forcing $F_1 = F_3$, but the following scaling allow to extensively generalise our findings: $q \sim (F_1)^{1/2}, V_{eff}(q) \sim (F_1)^{3/2}, \varepsilon_{eff} \sim (F_1)^{3/4}$.

Things change dramatically when considering the case of stochastic forcing acting only on one of the two high-latitude boxes. We tune our experiments in such a way that, apparently, the SDE for the $q$ variable is not altered with respect to the previous case. Not only we obtain a non-symmetric pdf, but, moreover, it is not possible to reconstruct an approximate but consistent stochastic dynamics for the $q$ variable alone. Therefore, there is no ground for achieving a satisfactory projection of the dynamics, and a one-dimensional Langevin equation cannot be derived.

Finally, we test the possibility of observing the mechanism of stochastic resonance in the simplified 2 d.o.f. model by superimposing a slow periodic modulation on the freshwater fluxes in the two high-latitude boxes to the acting stochastic forcings. Whereas in the scenario where the noise acts with equal strength in both boxes we obtain numerically outputs in close agreement with the theory stochastic resonance for one dimensional systems, thus supporting the idea that a projected dynamics is indeed a good approximation when attempting a description of the properties of $q$, the opposite holds in the scenario where the noise acts only on one box. This further supports the fact that in this case the dynamics of $q$ is not trivially quasi-one dimensional, and transitions occur through processes that cannot be written precisely as a function of $q$ only.

Our results support the idea that deducing the approximate stochastic dynamics for an observable of a multidimensional dynamical system from its time series is definitely a non-trivial operation. The reconstructed drift term and the noise forcing depend, in general, in a non-trivial way on the intensity and correlation properties of the white noise of the true system. This implies that a "true" projected dynamics cannot be defined. Therefore, in practical applications, it seems tentative to assume that from the pdf of a bimodal observable obtained with a given level of noise it is possible to understand how the bistability property of the full system will change when the level of the input noise is altered. In particular, it seems difficult to be confident on obtaining information on how the rate of transition between the two equilibria will change and on the characteristics of tipping points. A better understanding of the properties of multistable models can be reached only by going beyond a simplified description of the statistical properties of the observables we are mostly interested into. In order to address these points, we will attempt the kind of analysis proposed in this paper on more complex models of the thermohaline circulation, such as that considered in Lucarini et al. (2005, 2007).

## Acknowledgements

The authors wish to acknowledge the exchanges with J. Imbers and T. Kuna. VL and DF acknowledge the financial support of the EU-ERC project "Thermodynamics of the Climate System" – NAMASTE.

# Figures

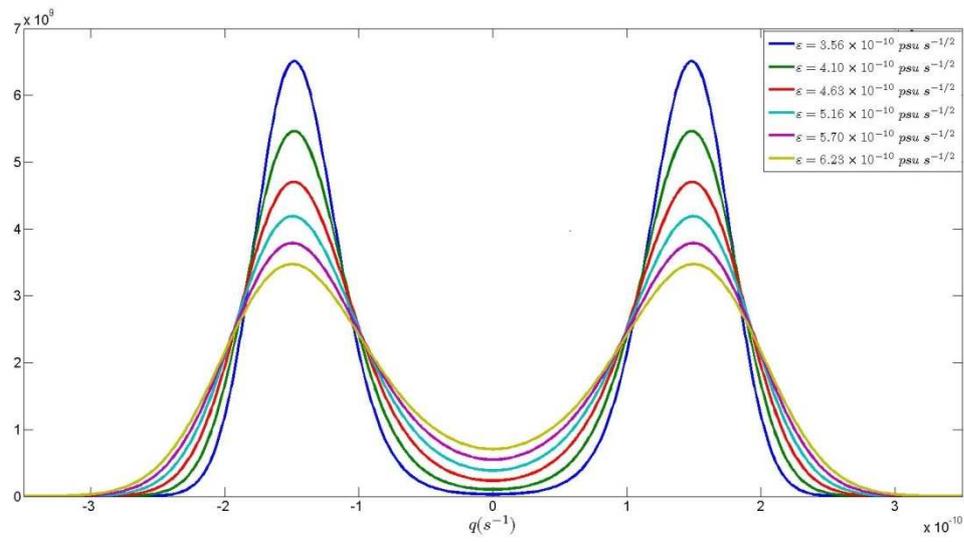

Figure 1: Empirical probability distribution function for the THC strength in the 2 d.o.f. model for selected values of symmetrically applied noise.

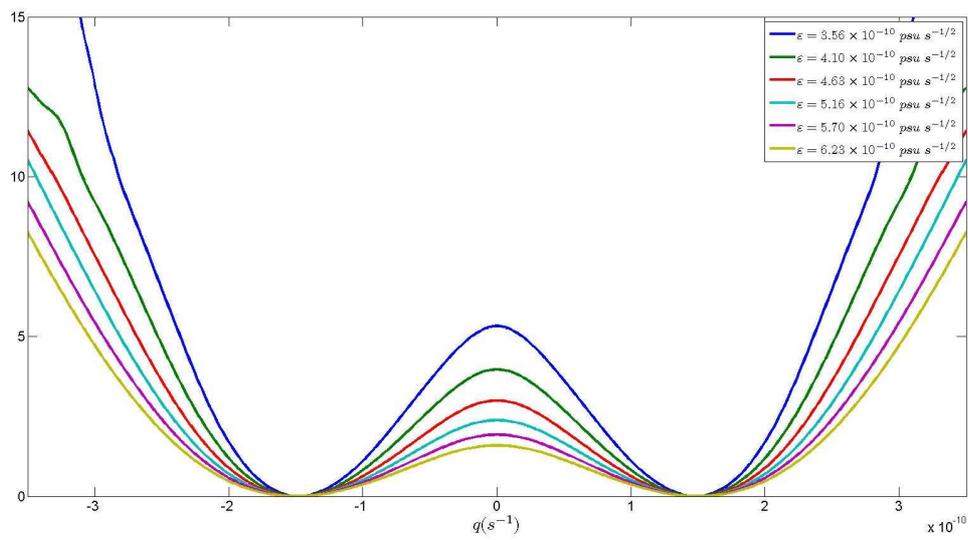

Figure 2: One-dimensional adimensional potential U(q) obtained as minus the logarithm of the pdf given in Fig. 1.

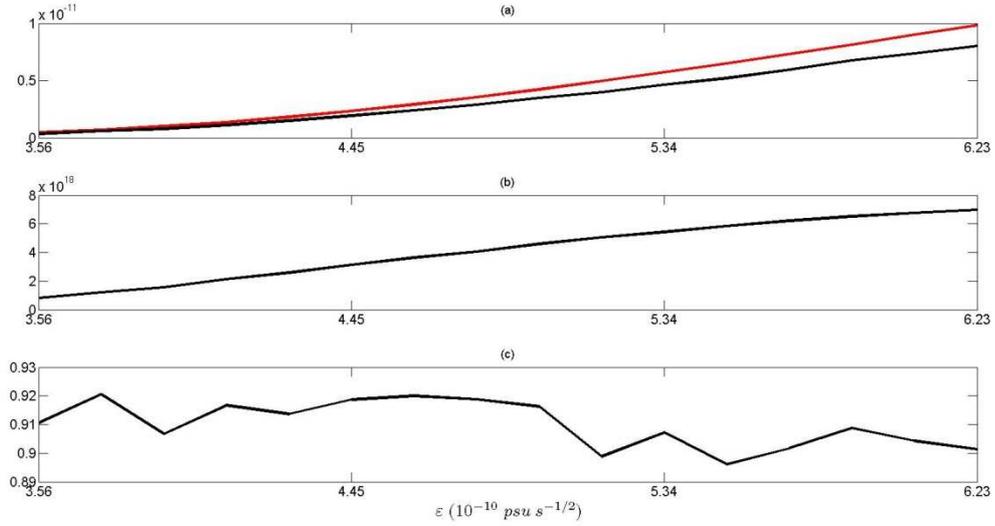

Figure 3: Goodness of the one-dimensional approach for the reduced 2 d.o.f. model. (a) Average rate of hopping between the northern and southern sinking equilibria r(+→-)=r(-→+) (black line). The theoretical value is shown with the red line. Data are in units of s$^{-1}$. (b) Geometrical factor G(+→-)=G(-→+) of the hopping rate computed from Eq. (7). Adimensional quantity. (c) Value of the parameter γ giving the ratio between $\varepsilon_{eff}$ (obtained as (a) divided by (b)) and the theoretically derived expression $\sqrt{2}k\beta\varepsilon$. Adimensional quantity.

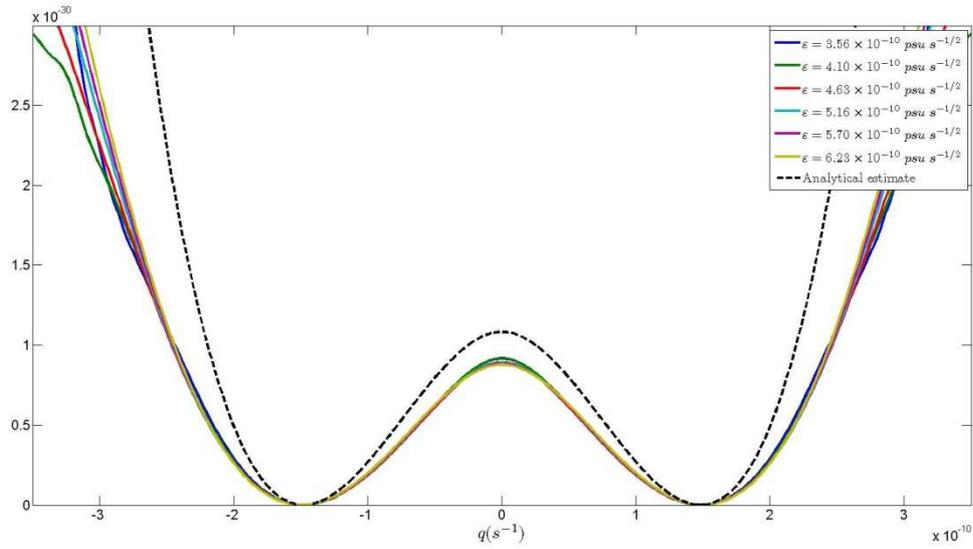

Figure 4: One-dimensional efficient potential V$_{eff}$(q) controlling the evolution of the THC strength in the 2 d.o.f. model with symmetric noise. The results of some numerical experiments are shown together with the analytical estimate.

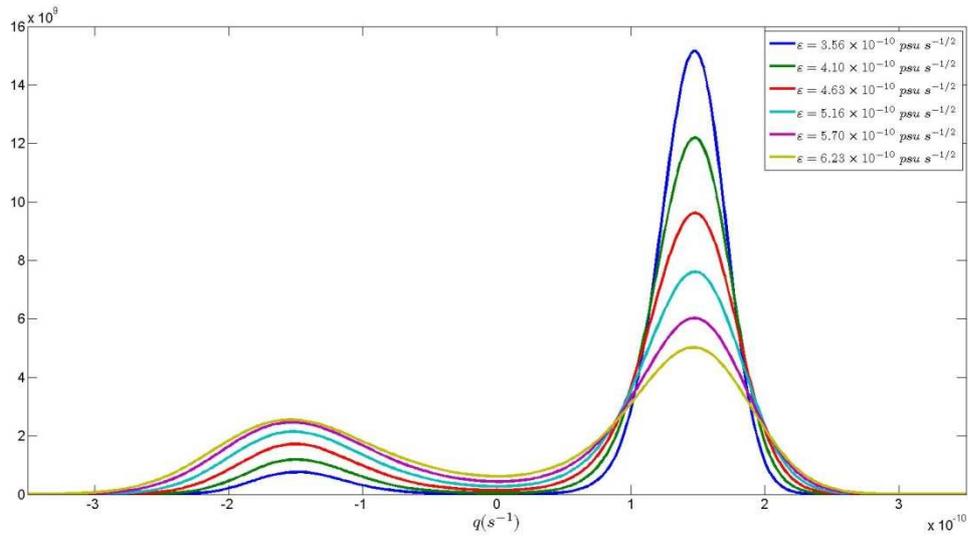

**Figure 5:** Empirical probability distribution function for the THC strength in the 2 d.o.f. model with noise applied only in box 1.

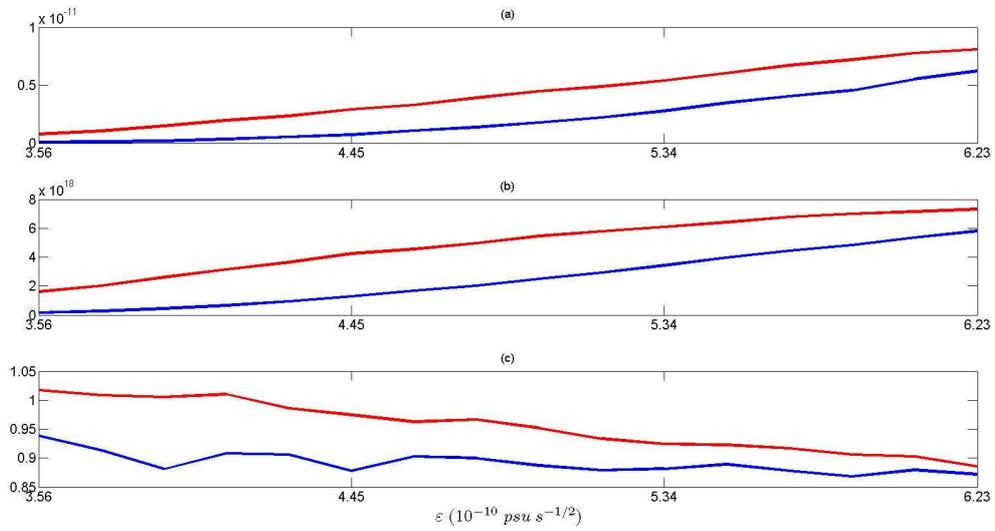

**Figure 6:** Goodness of the one-dimensional approach for the non-symmetric case. (a) Average hopping rates r(+→-) (blue line) and r(-→+) (red line). Data are in units of s$^{-1}$. (b) Geometrical factors G(+→-) (blue line) and G(-→+) (red line) computed from Eq. (7). Adimensional quantities. (c) Value of the parameter r γ giving the ratio between $\varepsilon_{eff}$ (obtained as (a) divided by (b)) and the theoretically derived expression $\sqrt{2}k\beta\varepsilon$. Adimensional quantities. The values obtained using the +→- (blue) and -→+ (red) processes are not compatible.

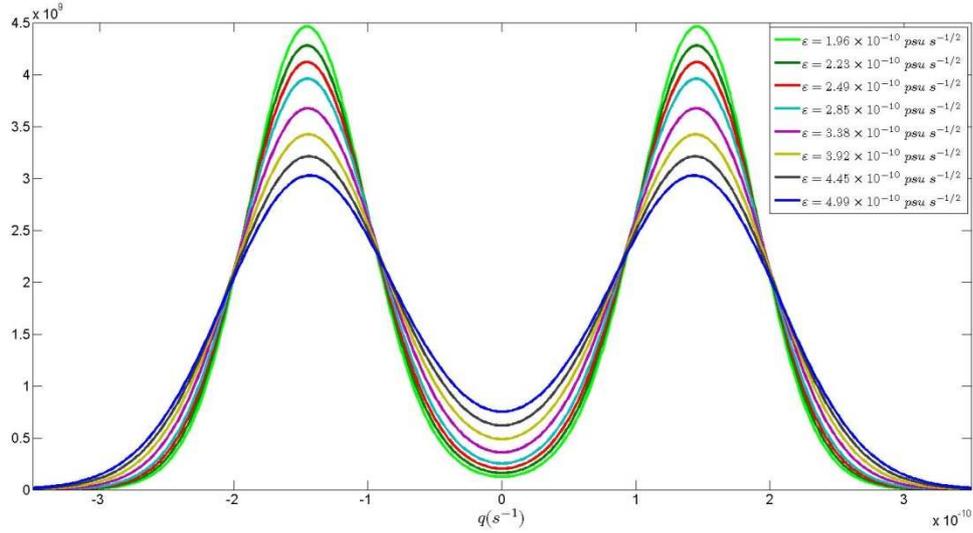

**Figure 7:** Empirical probability distribution function for the THC strength in the full model for some selected values of symmetrically applied noise.

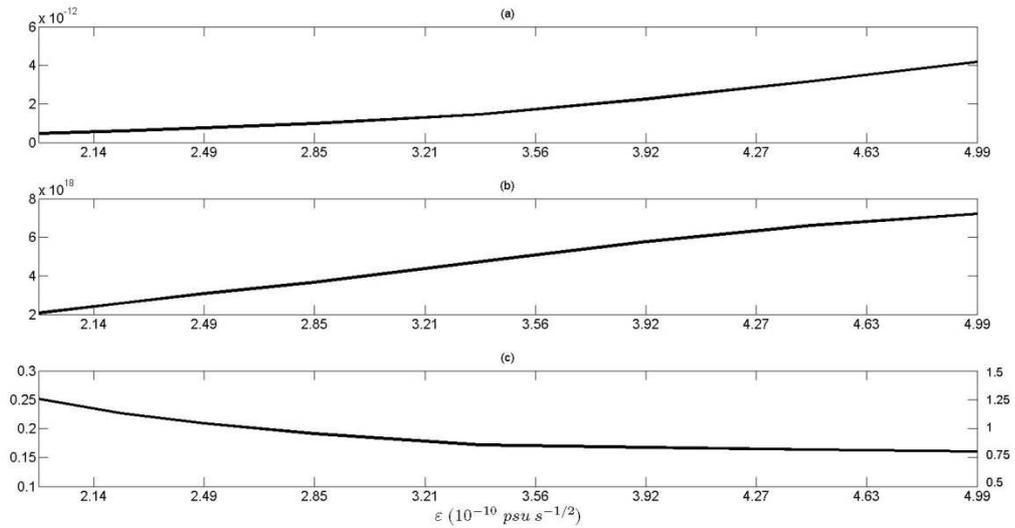

**Figure 8:** Goodness of the one-dimensional approach for the full model. (a) Average rate of hopping between the northern and southern sinking equilibria r(+→-)=r(-→+) (black line). The theoretical value is shown with the red line. Data are in units of s⁻¹. (b) Geometrical factor G(+→-)=G(-→+) of the hopping rate computed from Eq. (7). Adimensional quantity. (c) Value of the parameter γ giving the ratio between $\varepsilon_{eff}$ (obtained as (a) divided by (b)) and the theoretically derived expression $\sqrt{2}k\beta\varepsilon$. The value of γ obtained if considering the renormalized value (referred to the 2 d.o.f. model) for $k$ can be read on the right hand side scale. Adimensional quantities.

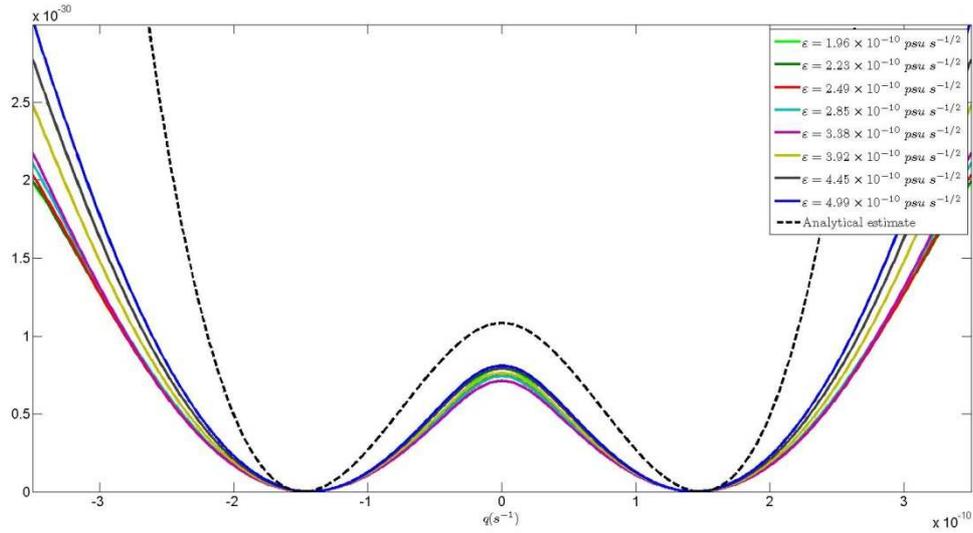

**Figure 9: One-dimensional efficient potential V$_{eff}$(q) controlling the evolution of the THC strength in the 5 d.o.f. model with symmetric noise. The results of some numerical experiments are shown together with the analytical estimate. The results are rather similar to what shown in Fig. 4.**

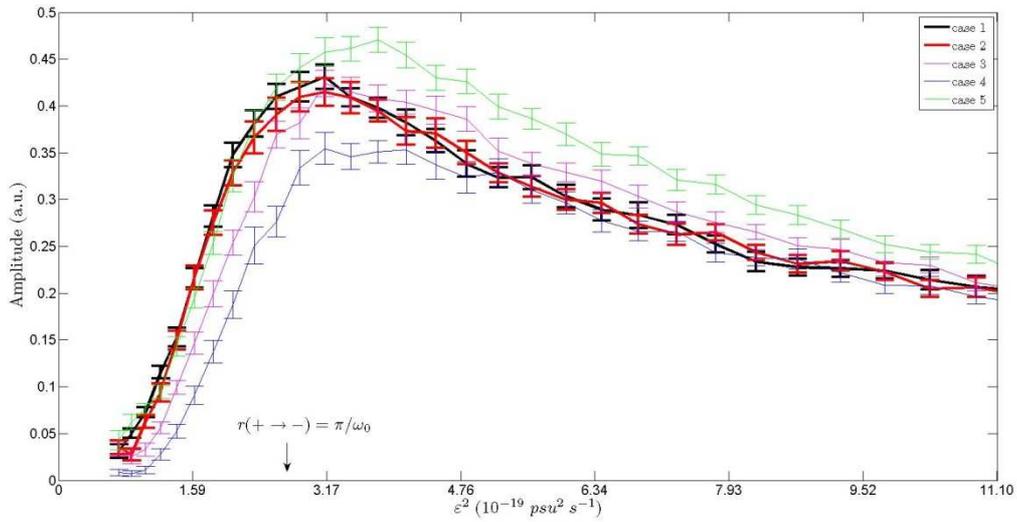

**Figure 10: Response to linear periodic perturbation as a function of the background noise for five experimental settings. Case 1 and 2 are indicative of stochastic resonance. The value of noise realising the approximate matching condition r(+→-)=r(-→+)=π/ω$_0$ is indicated.**